\documentclass[aps,prl,twocolumn,groupedaddress]{revtex4}
\usepackage[dvips]{graphicx,color}
\usepackage{amsmath,amssymb}

\begin{document}

\newcommand{\etal}      {{\it et~al.}}


\title{Search for $CP$-violation in Positronium Decay}


\author{T. Yamazaki, T. Namba, S. Asai, and T. Kobayashi}
\affiliation{Department of Physics, Graduate School of Science,
and International Center for Elementary Particle Physics, University of Tokyo, 
Hongo, Bunkyo-ku, Tokyo 113-0033, Japan}


\date{\today}

\begin{abstract}

$CP$-violation in the quark sector has been well established over the last decade,
but has not been observed in the lepton sector.
We search for $CP$-violating decay processes in positronium, 
using the angular correlation of
$(\vec{S}\cdot\vec{k_{1}})(\vec{S}\cdot\vec{k_{1}}\times\vec{k_{2}})$,
where $\vec{S}$ is the the positronium spin and $\vec{k_{1}}$, $\vec{k_{2}}$ are
the directions of the positronium decay photons.
To a sensitivity of $2.2\times10^{-3}$, no $CP$-violation has been found, which is
at the level of the $CP$-violation amplitude in the K meson. 
A 90\% confidence interval of the $CP$-violation parameter ($C_{CP}$) was determined to be $-0.0023 < C_{CP} < 0.0049$.  
This result is a factor 7 more strict than that of the previous experiment. 

\end{abstract}


\maketitle


The material state of the universe results only because of its matter-antimatter asymmetry.
To understand the origin of this asymmetry is one of particle physics' most urgent goals.
One cause could be the $CP$-violation (charge conjugation and parity) 
in the quark sector, which was found both in the K~\cite{kcp} and B~\cite{bcp1} mesons. 
The $CP$-violating processes in B mesons 
have been measured precisely~\cite{bcp2} using its various decay processes at KEKB and PEP-II
and they can be well described with the Kobayashi-Maskawa theory~\cite{KM}.
While understanding the origin of $CP$-violation in the quark sector
is a great achievement, the resulting asymmetry may be 
too small to result in the amount of matter in the Universe~\cite{cpx}.
All $CP$-violating observables in the Standard Model are proportional to the 
Jarlskog invariance, J$=s_1^2 s_2 s_3 c_1 c_2 c_3 \sin(\delta) \sim 3\times 10^{-5}$,
where $c_i$ are $s_i$ are the mixing angles in the KM matrix, and $\delta$ is the $CP$-violating phase.
The $CP$-violating observables are also suppressed by the Higgs vacuum expectation value, $1/v^{12}$, 
since non-equilibrium can occur in the electro-weak phase transition in the Standard Model. 
It seems that these violations alone are too small to accout for the observed baryons in the universe.
A new source of $CP$-violation or a new mechanism is necessary, 
and $CP$-violation in the lepton sector~\cite{yanagida} is a good candidate. 
Leptonic $CP$-violation has not yet been observed 
despite many studies at various neutrino factories~\cite{nufac}.

Positronium (Ps) is a neutral system comprising an electron and a positron and is the analogue of the neutral meson. 
If $CP$ is violated in the lepton sector, such neutral systems ({\it e.g.} Ps, muonium) are good test candidates since the admixture of opposite $CP$ eigenstates will occur.
The search for $CP$-violation in Ps has been performed previously
and no $CP$ violation was found at the 1.5\% level of uncertainty~\cite{skalsey}. 

If $CP$-symmetry is not conserved in Ps, the decay rate of the triplet Ps (o-Ps) is 
dependent on the $CP$-violating term.
The number of o-Ps decay events would be expressed as
\begin{equation}\label{eq:nEvent}
N = N_{0}[1+C_{CP}(\vec{S}\cdot\vec{k_{1}})(\vec{S}\cdot\vec{k_{1}}\times\vec{k_{2}})]\exp{(-t/\tau)},
\end{equation}
where $\vec{S}$ is the spin of the o-Ps, $\vec{k_{i}}$ is the direction of the $i$-th energetic 
$\gamma$-ray from the decay ($|\vec{k_{i}}|$=1), and $\tau$ is the lifetime of o-Ps.
$C_{CP}$ is the amplitude of the $CP$-violation.
The Standard Model prediction for $C_{CP}$ is of the order of $10^{-10}$~\cite{ma}. This is due to photon-photon interactions in the final state. 
Another exotic decay contribution has been excluded and should be
less than $2 \times 10^{-4}$~\cite{ex1,ex2,ex3,ex4,ex5}\@.
  
The discrete symmetry properties in the second term in Eq. (\ref{eq:nEvent}) are as follows:
All five vectors change sign under the $T$ operation,
while $\vec{k_{i}}$ and $\vec{S}$ are a vector and an axial vector, respectively,
under the $P$ operation.
Thus the angular correlation is $CP$-odd and $T$ odd 
but conserves $CPT$ symmetry.
This correlation would appear in general~\cite{ma} when the $CP$-violating effect exists in Ps.

The analyzing power $Q$ is defined as 
\begin{eqnarray}\label{eq:q}
Q     &=& (\vec{S}\cdot\vec{k_{1}})(\vec{S}\cdot\vec{k_{1}}\times\vec{k_{2}}) \nonumber \\
      &=& P_{2}\cdot\sin2\theta\sin\psi\cos\phi.
\end{eqnarray}
$\theta$ is the angle between the normal to the o-Ps decay plane 
and the spin quantization axis ($\vec{S}$) of the o-Ps, 
$\psi$ is the angle between $\vec{k_{1}}$ and $\vec{k_{2}}$, 
and $\phi$ is the angle between $\vec{k_{1}}$ and 
the projection of the spin quantization axis onto the o-Ps decay plane. 
Two parameters, $\theta$ and $\psi$, have been fixed during this measurement.
$C_{CP} Q(\phi)$ will be observed as an asymmetric function of $\phi$ 
if the $CP$-violation exists.
$P_{2}$ is the spin alignment of o-Ps (tensor polarization) and it can be expressed as
\begin{equation}\label{eq:p2}
P_{2} = \frac{N_{+1}-2N_{0}+N_{-1}}{N_{+1}+N_{0}+N_{-1}},
\end{equation}
where $N_{i}$ is the population of the $m_{s} = i$ Ps state. 
Usually the spin alignment $P_{2}$ is equal to zero 
since these three states are degenerate and thus their populations are 
expected to be the same, {\it i.e.} $N_{+1} = N_{0} = N_{-1}$.
A magnetic field is used to separate the triplet as follows: 
In a magnetic field, the o-Ps state with $m_{s} = 0$ and the p-Ps state mix,
and the resultant states have different energies (Zeeman effect).
This mixture depends on the strength of the magnetic field (0.2\% for 5 kG). 
The lifetime of the resultant o-Ps state with $m_{s} = 0$ is greatly reduced to about 22 ns 
although the lifetime of the perturbed singlet is almost the same as that of p-Ps (125 ps)~\cite{ppslife}.
On the other hand, the o-Ps states with $m_{s} = \pm1$ are not
affected at all in a static magnetic field
and their lifetimes are still 142.05 ns in vacuum~\cite{opslife1,opslife2}.
Since the $m_{s} = \pm1$ and $0$ states contribute opposite alignment 
as shown in Eq. (\ref{eq:p2}), we separate these Ps states on the basis of 
their different lifetimes in the external magnetic field. 


\begin{figure}
\includegraphics[width=85mm]{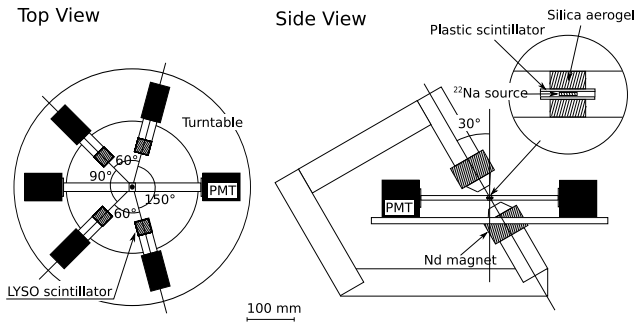}
\caption{Schematic diagrams of the experimental setup. 
The magnified view around the $^{22}$Na source is shown in the circle.
Left and right show the top and side view, respectively.\label{fig:setup}}
\end{figure}

Figure \ref{fig:setup} shows a schematic view of the experimental setup. 
A 1 MBq $^{22}$Na positron source is sandwiched 
between two sheets of thin plastic scintillator (NE102, t = 0.1 mm). 
The emitted positrons pass through the scintillator and produce 
light pulses that are directed to photomultipliers
(2-inch fine-mesh HAMAMATSU H6614-70) 
by the light guide shown horizontally in both top and side views of Fig. \ref{fig:setup}. 
The positrons form Ps when stopped in the targets of silica aerogel, 
10 mm (diam.) $\times$ 5 mm (thickness), and density 0.11 g/cm$^3$. 
In order to reduce the pick-off annihilation of the Ps we use aerogel in which the surface of the primary grain is made hydrophobic, and we also flush the aerogel with nitrogen gas (1 atm).

The gamma-rays emitted from Ps decay are observed in four LYSO (Lu$_{1.8}$Y$_{0.2}$SiO$_{5}$)
crystals (diameter = 30 mm and length = 30 mm) shown as the hatched boxes in the top view. 
The four detectors are placed such as to make three pairs 
with opening angles equal to 150$^{\circ}$.
These three pairs are used to detect the 3$\gamma$ decay of o-Ps.
The scintillation pulses of the LYSO crystals are detected 
with $1\frac{1}{2}$-inch fine-mesh photomultipliers (HAMAMATSU H8409-70).
The average energy resolution of the LYSO detectors is 60 keV (FWHM) at 511 keV, and 
a good timing resolution of 1.2 ns (FWHM) is obtained for 511 keV gamma-rays. 
Since the mean atomic number (Z) of the LYSO crystal is large, 
the photoelectric process dominates compared to the Compton scatter process.
This is an advantage for tagging $\vec{k_{1}}$ and $\vec{k_{2}}$ correctly. 

A magnetic field to perturb the Ps spin states is generated using two 80 mm (diam.) 
$\times$ 50 mm (thickness) Nd permanent magnets with iron cones and yokes as shown in the side view. 
The magnetic field is 4.9 kG and is directed at 30$^{\circ}$ 
from the direction of the normal to the 3$\gamma$ decay plane.
Its uniformity is about 10$\%$ over the volume of the silica aerogel.
The observed unperturbed and perturbed o-Ps lifetimes are 126 ns and 22.5 ns, 
respectively. These lifetimes are consistent with the expected lifetimes 
in 1 atm nitrogen gas and a magnetic field of 4.9 kG.

As shown in the top view, all scintillators are 
located on a turntable which turns with respect to 
the magnetic field.
Using this turntable, we can vary $\phi$ in Eq. (\ref{eq:q}) independently 
and measure the $\phi$-dependence of the number of events.
The turntable is rotated by a stepping motor and its rotation angle is 
controlled using a CAMAC system. 
The angle of the turntable is measured by an optical sensor 
with 0.2$^{\circ}$ accuracy. 
The experimental apparatus is carefully assembled to avoid geometrical 
asymmetry and the assembly accuracy is about 0.5 mm.

In order to select Ps decay events data acquisition logic is set up
as follows:
When at least two signals from the LYSO scintillators are coincident within 20 ns, 
and then when this coincidence is within $-$100 ns to 600 ns 
of the timing of the plastic scintillators, data aqcuisition is triggered.
The coincidence between the LYSO detectors reduces accidental counts 
due to the radioactive decay of $^{176}$Lu in the crystal.
A charge ADC (REPIC RPC-022) is used to measure the energy information of 
the plastic scintillators. The outputs of the LYSO detectors are recorded 
with another charge ADC (CAEN C1205).
The time information of the plastic and LYSO scintillators is recorded 
using a direct clock (2 GHz) count type TDC~\cite{opslife1}.

Five runs with different alignments of the LYSO detectors were 
performed to check systematic effects dependent on the position of the LYSO 
detectors. The total data acquisition period was about six months. 
The trigger rates were about 1.3 kHz.
During the data acquisition, the turntable was rotated around 30$^{\circ}$ 
every hour. Energy and time calibrations were also performed 
every hour. The $\gamma$-ray peak at 511 keV and the zero energy peak were used 
to calibrate the LYSO detectors. 
The room temperature was maintained within $22 \pm 1 ^{\circ}$C in order to maintain good 
stability during the data acquisition. 

\begin{figure}
\includegraphics[width=65mm]{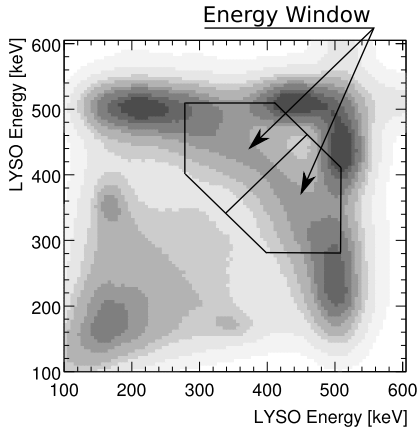}
\caption{2D energy distribution of o-Ps decay $\gamma$'s obtained by 150$^{\circ}$ pairs of the LYSO detectors.\label{fig:edep_LYSOs}}
\end{figure}

Figure \ref{fig:edep_LYSOs} shows the 2-dimensional distribution of $\gamma$- 
energies obtained with the 150$^{\circ}$-pairs of the LYSO detectors. 
$\gamma$-rays from $3\gamma$ decays are distributed in the arc whose shape is
determined by the opening angle of the detector pairs.
To select 3$\gamma$ events, the following energy-window is applied:
\begin{gather}
680\ {\rm [keV]} < E_{1} + E_{2} < 920\ {\rm [keV]}, \\
280\ {\rm [keV]} < E_{2} < E_{1} < 511\ {\rm [keV]},
\end{gather}
where $E_{i}$ is energy of $i$-th energetic $\gamma$-ray emitted from the o-Ps decay. 
This selection is superimposed in Fig. \ref{fig:edep_LYSOs}. 

\begin{figure}
\includegraphics[width=85mm]{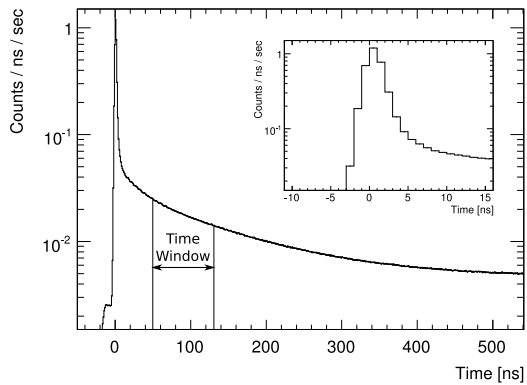}
\caption{Time difference between the plastic scintillators and 
the coincidence signals of the LYSO scintillators
after the energy selection is applied. 
The figure in the right corner is
the magnified view of the sharp peak of the prompt annihilation.\label{fig:lifetime}}
\end{figure}

Figure \ref{fig:lifetime} shows the time difference between the 
plastic scintillator signal and the coincidence signals of the LYSO detectors 
after the energy selection is applied.
A sharp peak from prompt annihilation is followed by the exponential curves of 
o-Ps and then the constant accidental spectrum. 
The figure in the right corner shows the magnified 
view of the sharp peak of the prompt annihilation. 
A good timing resolution of $\sigma = 0.9$ ns is obtained.
A time window from 50 ns to 130 ns is required to 
enhance the $m_{s} = \pm1$ Ps states and a spin alignment $P_2$ = 0.87 is obtained.

\begin{figure}
\includegraphics[width=85mm]{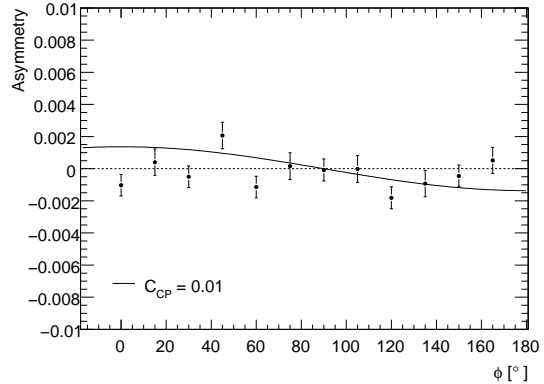}
\caption{Measured asymmetry averaged over the detector pairs and five runs. 
The line shows the expected angular dependence 
if the $CP$ violating coefficient $C_{CP} = 0.01$.\label{fig:asymmetry}}
\end{figure}

The asymmetry, $A(\phi)$, is calculated using the numbers of events, $N(\phi)$,
from which the accidental contribution estimated with the time spectrum shown in Fig. \ref{fig:lifetime} is subtracted:
\begin{gather}
A(\phi) = \frac{N(\phi)-N(\phi+180^{\circ})}{N(\phi)+N(\phi+180^{\circ})}.
\end{gather}
This asymmetry is equal to $C_{CP}Q(\phi)$ since the numbers of events are 
expressed as Eq. (\ref{eq:nEvent}) and the analyzing power $Q(\phi)$ depends on 
$\cos\phi$ as shown in Eq. (\ref{eq:q}). 
The differences between the LYSO detector pairs do not affect the measured asymmetry since an independent asymmetry measurement is performed for each detector pair (thanks to the rotating table).
The effect of Compton scatter by the magnet near the $^{22}$Na source is
obviously the same between $N(\phi)$ and $N(\phi+180^{\circ})$ 
since the magnet system is symmetric with respect to the origin of the 
experimental setup.
In addition, averaging the asymmetries of the same analyzing power over the 
pairs of the LYSO detectors cancels out systematic effects due to other 
unknown sources.
In Fig. \ref{fig:asymmetry}, the asymmetries are plotted as a function 
of $\phi$.
If $CP$-symmetry is violated, the $A(\phi)$ distribution would be a cosine 
curve (superimposed in the figure) whose amplitude is proportional to 
the magnitude of the $CP$-violation.

The obtained data is consistent with the 
flat distribution as shown in Fig. \ref{fig:asymmetry} 
and thus no $CP$-violating effect is observed. 
The $C_{CP}$ coefficient is obtained by dividing the measured asymmetries by the analyzing power. 
The analyzing power shown in Eq. (2) is the ideal case, but 
the real analyzing power of this setup is affected by various factors: 
the finite solid angle ($-$15\% with respect to the ideal case), 
the energy resolution ($-$2\%) of the LYSO crystals, 
and the finite size of the Ps creation position ($-$1\%).
These were estimated using a Monte Carlo simulation.
The angular correlation of the $\gamma$-rays  from the $m_{s} = 0, \pm1$ states of o-Ps 
is also taken into account using a theoretical calculation~\cite{weak}. 
The obtained coefficient averaged over $\phi$ is 
$C_{CP} = 0.0013 \pm 0.0021(stat.)$,
and $C_{CP}$ is consistent with zero within the statistical
uncertainty. 

Estimates of various systematic errors are summarized in Table \ref{tab:sys}.
The largest contribution is the uncertainty in the angle of the turntable.
The turntable angle is directly related to $\phi$,
and the effect of Compton scatter by the magnet can not be canceled out 
if the angle between $N(\phi)$ and $N(\phi+180^{\circ})$ differs from 180$^{\circ}$. 
The number of events differed by $\pm$10\% when the table was rotated by 90$^{\circ}$
due to this geometrical effect and 
this difference is applied to the 180$^{\circ}$-case conservatively. 
An uncertainty (0.2$^{\circ}$) in the table angle could contribute an 
uncertainty of about $2\times 10^{-4}$ to the event number. 
This error propageted to $C_{CP}$ is $3.9\times 10^{-4}$.
Another predominant systematic error is produced if the center of the turntable 
and the magnetic field are misaligned. 
In such cases, the number of events which pass the time window changes 
since the mean magnetic field over the Ps creation area depends on 
the table angle.
From the measured magnetic field, and the assembly accuracy of the experimental 
setup (0.5 mm), we can estimate the uncertainty of the mean magnetic field, and 
it contributes to the error on $C_{CP}$ by $2.5\times 10^{-4}$.
Non-uniformities in the scilica aerogel and the magnetic field would
also produce asymmetries, 
since they also make the effective strength of the magnetic field depend on the table angle.
The last predominant source is the decrease in the decay rate of $^{22}$Na 
($T_{1/2} = 2.6$ yr).
In this experiment, $N(\phi)$ and $N(\phi+180^{\circ})$ are recorded 
at different times (time difference $= 6$ h). The decay rate 
of $^{22}$Na decreases slightly over this time ($2.6\times 10^{-4}$).
This results in a small decrease in the number of events and causes an asymmetry.

\begin{table}
\caption{Summary of the systematic errors.\label{tab:sys}}
\begin{ruledtabular}
\begin{tabular}{lc}
\multicolumn{1}{c}{Source}                 & Systematic error      \\
\hline
Table angle accuracy                       & $\pm 0.00039$         \\
Center alignment                           & $\pm 0.00025$         \\
Non-uniformity of aerogel / magnetic field & $\pm 0.00011$         \\
Decrease of $\beta^{+}$ decay rate         & $\pm 0.00030$         \\
\end{tabular}
\end{ruledtabular}
\end{table}

The systematic errors discussed above are regarded as independent contributions 
such that the total systematic error can be calculated as their quadratic sum. 
The combined result with the systematic errors is 
\begin{equation}
C_{CP} = 0.0013 \pm 0.0021(stat.) \pm 0.0006(sys.).
\end{equation}
$C_{CP}$ is consistent with zero and a 90\% confidence interval of 
\begin{equation}
-0.0023 < C_{CP} < 0.0049,
\end{equation}
is obtained.
This result is 7 times more strict than that of the previous experiment~\cite{skalsey}
and is of a similar sensitivity to that of the Kaon $CP$-violation.

In order to perform more accurate experiment you need more statistics but you can not simply use a radioactive source with higher intensities because it causes pulse pile-up in detector systems. 
A possible way to increase statistics is to use detector arrays with nearly 4$\pi$ sr of coverage. 
In addition, as for systematics, an electromagnet can be useful because it can also cancel many systematic effects by reversing its polarity.   

A $CP$-violating decay process in positronium
has been searched for using the angular correlation of
$(\vec{S}\cdot\vec{k_{1}})(\vec{S}\cdot\vec{k_{1}}\times\vec{k_{2}})$,
where $\vec{S}$ is the the positronium spin and $\vec{k_{1}}$, $\vec{k_{2}}$ are the directions of the decay photons.
LYSO scintillators are used to measure the photon energies
and these are set up to turn with respect to the magnetic field in
order to cancel out various systematic errors.
To a sensitivity of $2.2\times10^{-3}$, no $CP$-violation has been observed. 
This level is similar to the $CP$-violation amplitude of the K meson. 

We thank Prof. K. Hamaguchi (U. Tokyo) and Prof. T. Moroi (U. Tohoku) 
for theoretical discussions on $CP$-violation.
Sincere gratitude is expressed to Mr. M. M. Hashimoto 
and Mr. K. Nishihara for useful discussions.
This research was funded in part by the Japan Society for the Promotion of Science.

\end{document}